\documentclass[conference]{IEEEtran}
\IEEEoverridecommandlockouts
\usepackage{cite}
\usepackage{amsmath,amssymb,amsfonts}
\usepackage{algorithmic}
\usepackage{graphicx}
 \usepackage{lettrine}
\usepackage{textcomp}
\usepackage{xcolor}
\def\BibTeX{{\rm B\kern-.05em{\sc i\kern-.025em b}\kern-.08em
    T\kern-.1667em\lower.7ex\hbox{E}\kern-.125emX}}
\usepackage{amsmath}
\usepackage{amssymb}
\usepackage{amsthm}
\usepackage{amsfonts}
\usepackage{hyphenat}
\usepackage{hyperref}
\usepackage{times}
\usepackage{psfrag}
\usepackage{ifpdf}
\usepackage{array}
\usepackage{multirow}
\usepackage{setspace}
\usepackage{latexsym}
\usepackage{enumerate}
\usepackage{mathtools}
\usepackage{float}
\usepackage{scalerel}
\usepackage{cite}
\usepackage{cleveref}
\usepackage{graphicx}
\usepackage{caption}
\usepackage{epstopdf}
\usepackage{subfig}
\usepackage{accents}
\usepackage{algorithm}
\usepackage{algorithmic} 
\usepackage{mathrsfs}
\usepackage{tabularx}
\usepackage{tikz}
\usepackage{pgfplots}
\usepackage{url}  %Required
\usepackage{caption}
\usepackage{rotating}

\definecolor{Fcolor}{rgb}{0,0.5,0.25}

\theoremstyle{definition}

\crefrangelabelformat{equation}{(#3#1#4)\,--\,(#5#2#6)}
\crefmultiformat{equation}{(#2#1#3)}{\,--\,(#2#1#3)}{#2#1#3}{#2#1#3}

\crefname{equation}{}{}
\crefname{figure}{Figure}{Figures}
\crefname{algorithm}{Algorithm}{Algorithms}
\crefname{table}{Table}{Tables}
\crefname{lemma}{Lemma}{Lemmas}
\crefname{theorem}{Theorem}{Theorems}
\crefname{remark}{Remark}{Remarks}
\crefname{section}{Section}{Sections}
\crefname{subsection}{Subsection}{Subsections}
\crefname{definition}{Definition}{Definitions}

\newcommand{\Abf}{\boldsymbol{A}}

\newcommand{\Bbf}{\boldsymbol{B}}

\newcommand{\Cbf}{\boldsymbol{C}}

\newcommand{\Ccal}{\mathcal{C}}

\newcommand{\dbf}{\boldsymbol{d}}

\newcommand{\Ecal}{\mathcal{E}}

\newcommand{\fbf}{\boldsymbol{f}}

\newcommand{\Gcal}{\mathcal{G}}

\newcommand{\pbf}{\boldsymbol{p}}

\newcommand{\Rbb}{\mathbb{R}}

\newcommand{\Tcal}{\mathcal{T}}

\newcommand{\Vcal}{\mathcal{V}}

\makeatletter
\DeclareRobustCommand{\cev}[1]{%
  \mathpalette\do@cev{#1}%
}
\newcommand{\do@cev}[2]{%
  \fix@cev{#1}{+}%
  \reflectbox{$\m@th#1\vec{\reflectbox{$\fix@cev{#1}{-}\m@th#1#2\fix@cev{#1}{+}$}}$}%
  \fix@cev{#1}{-}%
}
\newcommand{\fix@cev}[2]{%
  \ifx#1\displaystyle
    \mkern#2 1mu
  \else
    \ifx#1\textstyle
      \mkern#2 3mu
    \else
      \ifx#1\scriptstyle
        \mkern#2 2mu
      \else
        \mkern#2 2mu
      \fi
    \fi
  \fi
}

\begin{document}

\title{Proactive Posturing of Large Power Grid for Mitigating Hurricane Impacts\\
\vspace{-1ex}
{\footnotesize \textsuperscript{}}
\thanks{This work is supported by the U.S. Department of Energy and Federal Emergency Management Agency. Pacific Northwest National Laboratory (PNNL) is operated by Battelle for the DOE under Contract DOE-AC05-76RL01830.}
}

\author{\IEEEauthorblockN{Edward Quarm Jnr., Xiaoyuan Fan, Marcelo Elizondo}
\IEEEauthorblockA{\textit{Energy and Environment Directorate} \\
Pacific Northwest National Laboratory \\
Richland, WA USA \\
\{edwardarthur.quarmjnr; xiaoyuan.fan; marcelo.elizondo\}@pnnl.gov
\vspace{-2ex}}
\and
\IEEEauthorblockN{Ramtin Madani}
\IEEEauthorblockA{\textit{Electrical Engineering Department} \\
University of Texas at Arlington \\
Arlington, TX USA \\
ramtin.madani@uta.edu
\vspace{-2ex}}

% \author{\IEEEauthorblockN{Edward Quarm Jnr.}
% \IEEEauthorblockA{\textit{Pacific Northwest National Laboratory} \\
% Richland, WA USA \\
% xiaoyuan.fan@pnnl.gov}
% \and
% \IEEEauthorblockN{Xiaoyuan Fan}
% \IEEEauthorblockA{\textit{Pacific Northwest National Laboratory} \\
% Richland, WA USA \\
% xiaoyuan.fan@pnnl.gov}
% \and
% \IEEEauthorblockN{Marcelo Elizondo}
% \IEEEauthorblockA{\textit{Pacific Northwest National Laboratory} \\
% Richland, WA USA \\
% xiaoyuan.fan@pnnl.gov}
% \and
% \IEEEauthorblockN{Ramtin Madani}
% \IEEEauthorblockA{\textit{University of Texas at Arlington} \\
% Arlington, Texas USA \\
% ramtin.madani@uta.edu}

}

\maketitle
\begin{abstract}
In the past decade, natural disasters such as hurricanes have challenged the operation and control of U.S. power grid. It is crucial to develop proactive strategies to assist grid operators for better emergency response and minimized electricity service interruptions; the better the grid may be preserved, the faster the grid can be restored. In this paper, we propose a proactive posturing of power system elements, and \textcolor{black}{formulate} a Security-Constrained Optimal Power Flow (SCOPF) informed by cross-domain hurricane modeling as well as its potential impacts on grid elements. Simulation results based on real-world power grid and historical hurricane event \textcolor{black}{have verified} the applicability of the proposed optimization formulation, \textcolor{black}{which shows} potential to enable grid operators and planners with interactive cross-domain data analytics for  mitigating  hurricane impacts.%show improvements in decreased xxx and increased yyy, as well as zzz.
\end{abstract}

\begin{IEEEkeywords}
Power system emergency response, security constrained optimal power flow, hurricane impact mitigation
\end{IEEEkeywords}
%\section*{Nomenclature}
%\input{Nomenclature}

\section{Introduction}

% Introduction =====================================
\lettrine{S}evere weather is one of the major threats to power system security in the United States today. The effects of climate change and worsening pollution levels have contributed to destructive weather events such as hurricanes, tornadoes, blizzards, and droughts \cite{BillionDollarEvents}.
%Therefore, it is crucial to develop a framework for establishing critical infrastructure resilience goals \cite{NIACReport2021}, and using it as one important guideline for multi-domain, multi-agency coordination, planning, and emergency response.
For example, in 2017, Hurricane Irma and Maria stroke Puerto Rico as Category 5 storm, and caused prolonged yet wide-spread damage to Puerto Rico's electrical infrastructure; \textcolor{black}{and in 2021, Hurricane Ida became the second-most damaging hurricane for Louisiana on record}. Research efforts have been made to develop new algorithms and tools in support of optimized grid resilience improvements and potential emergency  mitigation strategies \cite{osti_1771798,osti_1764377}. \textcolor{black}{More importantly, industry guidelines and comprehensive evaluation framework should be properly defined to achieve critical infrastructure resilience goals\cite{NIACReport2021} and enable multi-domain, multi-agency coordination, planning, and emergency response.}

% Literature Review =====================================

Several research papers have explored different ways of including hurricane contingencies into the optimal power flow (OPF) and unit commitment (UC) problems. Current modeling approaches can be widely classified into two categories: proactive operation, and corrective operation of the grid in response to contingency events. Studies \cite{corrective1,corrective2,corrective3} explored corrective approaches to contingency modeling where  \cite{preventive1,preventive2,preventive3} explored proactive approaches. %Real-time Corrective transmission switching (RTCTS) was studied in \cite{corrective2} as a viable corrective action for handling contingencies, however concerns regarding stability  and computational complexity makes practical implementation challenging.
Proactive modeling strategies such as \cite{preventive3} has shown some promise in mitigating the adverse impacts of contingencies on the grid. %In \cite{preventive3} an iterative process is proposed to generate mitigation cuts which are then incorporated into an enhanced Security-constrained Economic Dispatch (SCED) problem. %In \cite{preventive4}, contingencies were filtered by discarding those which led to fewer violations in a preventive Security-Constrained Optimal Power Flow (SCOPF) problem. However, in most research papers the network size considered is smaller than most real world grids.

Large-scale SCOPF problems are computationally intractable \cite{NPhard3,NPhard4}. The complexity is attributed to the presence of nonlinear alternating current (AC) network constraints, enforcing large number of transmission constraints, considering multiple scenarios and multiple contingencies in the formulation\cite{NPhard1,NPhard2}. In order to alleviate the computational burden in $N-k$ OPF problems, research papers \cite{NPhard1, screening1}  proposed contingency screening to reduce problem size.% algorithm to identify contingencies that are critical in DC approximation. In , authors reported success in handling larger systems after implementing transmission line contingency screening algorithm.

Semi-definite programming (SDP) relaxation holds significant promise for application to large-scale OPF and UC problems. In \cite{SDP1}, the authors proved that a global optimum solution to the OPF problem can be obtained when the duality gap is zero. However, for some practical problems SDP relaxation may not be tight, %therefore,
\textcolor{black}{and it} fails to obtain a global optimum solution. In order to address inexactness of SDP relaxation, penalty terms are incorporated  into  the  objective  of  convex  relaxations  to drive the solution to near-globally optimal solution \cite{SDP2,NPhard2,SDP3,SDP4}. Furthermore, \cite{SDP5} proposed adding valid inequalities to strengthen SDP relaxation when the relaxation is inexact.

%Second-order Cone Programming (SOCP) relaxations are computationally cheaper relaxations as compared to SDP relaxations. SOCP relaxations have been applied in \cite{socp1,socp2}. Some papers have leveraged the sparsity of power networks to decompose large-scale conic constraints into lower order ones \cite{NPhard2,socp3} by applying chordal decomposition.

% Contributions =====================================
\textcolor{black}{Limited} %Very few
papers have attempted to consider the sequential temporal relationship between individual \textcolor{black}{hurricane} contingencies in the multi-period SCOPF problem. \textcolor{black}{For example, \cite{ZZLYW2021} presented a two-stage unit commitment optimization formulation considering equipment linear failure probability due to hurricane impacts, but missed the detailed grid responses based on electromechanical and protection behaviors.} In this paper, we adopt a proactive strategy of handling contingencies brought about by hurricane events with the aim of improving the preparedness of the grid to withstand the disruptive events, \textcolor{black}{in which both optimization tools and grid time-domain simulation tools are utilized}. The temporal relations among  individual  steps  (groups  of  time  period)  within  one  hurricane  event  have  been  explored, in particular the present group of power system contingencies (includes  outaged generators and transmission line) will be considered as N-1 security constraints, while the following group of power system contingencies is incorporated as explicit line rating constraints; this cascaded forward-looking design embodies proactive  posturing  of  power  system  elements, but differentiates itself from simply stacking adjacent groups of power system contingencies. To tackle this, We employ a low order conic relaxation which has been demonstrated through multiple experiments in \cite{21QEMR} to be exact for many large-scale SCUC problem and suitable to be applied to SCOPF formulations.

% Structure of paper =====================================
The rest of this paper is organized as follows: section \ref{hurricane_model_optimization} introduces modeling hurricane's impact on the electric grid. Next, the SCOPF with contingencies is formulated in \ref{scopf_formulation}. In section \ref{convexification}, we describe a tractable convex relaxation for the optimization problem by means of convex surrogates. Next, a semidefinite programming relaxation is proposed to tackle the problem. Numerical simulations are provided in section \ref{numerical_results} and conclusions are summarized in section \ref{conclusions}.

% Noatations =====================================
\subsection{Notations}
Throughout this paper, matrices, vectors and scalars are represented by boldface uppercase, boldface lowercase and italic lowercase letters, respectively.  $|\cdot|$ represents the absolute value of a scalar or the cardinality of a set. The symbol $(\cdot)^\top$ represents the transpose operator. The notation real$\{\cdot\}$ represents the real part of a scalar or a matrix. Given a matrix $\Abf$, the notation $\Abf$, the notation $\Abf_{jk}$ refers to its $(j,k)^{th}$ element. $\Abf$, he notation $\Abf_{jk}$ refers to its $(j,k)^{th}$ element. $\Abf\succeq 0$ means that $\Abf$ is symmetric and positive semidefinite.

\section{Modeling Hurricane's Impacts on Grid and the Optimization Problem Formulation}\label{hurricane_model_optimization}
\textcolor{black}{Hurricane is} one of the major natural disasters that influence U.S. every year, the modeling and prediction of hurricane based on climatology and meteorology are critical and fundamental \cite{e3sm-model}. But to fully capture its impacts on power grid and further take advantage of such cross-domain capabilities into power grid operation and planning, major barriers including GIS data fusion and standardized cross-domain data curation pose significant challenges for private industry \cite{osti_1764377}, not even mention the following requirements on optimization and computing techniques when developing system hardening strategies and impact mitigation plan.

To overcome those challenges, we formulate a multi-period SCOPF problem, which considers hurricane-influenced grid contingencies as well as the sequential temporal relationship between individual steps. As a result, the modeling and prediction of hurricane, usually published 24 to 72 hours in advance, can be transformed into cross-domain intelligence in a proactive manner, and further incorporated into power system domain analysis. More details are given as follows.
\subsection{Hurricane Contingency Description}\label{hurricane_description}
Hurricane can be modelled as a multi-step temporal process, each step includes the damaged group of power system equipment and may be described as classical contingency for power grid, such as line tripping, generation tripping, etc.. It should be noted that individual power system equipment has its own fragility curve, which dictates the failure possibility of that equipment under certain wind speed during hurricane or storm \cite{osti_1771798}. For example, one performed simulation for Hurricane Maria included three steps, and each step includes one PSS\textregistered E idv file to represent the sequence of elements damaged by hurricane. Therefore, each idv file is considered as a form of dynamic contingency simulation with proper time spacing between individual equipment failure, while it is also natural to be considered as $N-k$ contingency formulation in steady-sate analysis.

A preventive SCOPF problem can be formulated for each step, by taking advantage of temporal relationship among the cascaded hurricane contingency groups and maximize the benefits of the predictable hurricane trajectory and intensity in advance. More specifically, by proactively posturing the power system operating conditions to embrace the possible grid contingencies due to power system equipment failure from both temporal and geographical perspectives, the proposed method aims to explore the optimal system operation point considering the current group of contingencies as security constraints, while also minimizing the potential impacts of the following batch of contingencies by penalizing line flows on candidate lines accordingly. As a result, the cross-domain intelligence from advanced hurricane modeling and prediction can be transformed into timely absorbable information to guide proactive posturing of power grid.

\subsection{Multi-period Proactive SCOPF Problem Formulation}\label{scopf_formulation}
Let $\mathcal{T}$ represent the set of discrete time slots, with index $t$. Additionally, let $\Gcal_{t}$ represent the set of all generating units that are available at time $t\in \Tcal$. Define $\Ccal_{t}$ as the set of all predicted hurricane contingency cases with $0\in\Ccal_{t}$ representing the base case (normal grid operation). Lastly, $\Gcal_{tc}\subseteq\Gcal_{t}$ is defined as the set of all generating units that are available at time $t\in\Tcal$ and contingency $c\in\Ccal_{t}$, such that
\begin{align}
\Gcal_{t}= \cup_{c\in\Ccal_{t}}\Gcal_{tc}.\nonumber
\end{align}

Consider a power system with $\Vcal$ as the set of buses, and $\Ecal_{tc}\subseteq\Vcal\times\Vcal$ as the set of branches at time $t\in \Tcal$ and contingency $c\in\Ccal_{t}$. The proactive SCOPF problem is formulated as:
%============================================================

%============================================================
\begingroup
\allowdisplaybreaks
\begin{subequations}\label{initial_problem}
\begin{align}
&\underset{}{\text{min~~~}} && \!\! \sum_{t \in \mathcal{T}}^{} \Bigg(
\sum_{c \in \mathcal{C}_{t}}^{}
\sum_{g \in \mathcal{G}_{tc}}^{}
\alpha^{\mathrm{sqr}}_{tg}\; p^2_{tgc} +\alpha^{\mathrm{lin}}_{tg}\; p_{tgc} \,+ \zeta_{tg} \label{c1a}\\
& && + \sum_{c \in 0}^{}
\sum_{g \in \mathcal{G}_{tc}}^{} \kappa_{g}\!\times\!(p_{tg0} - p_{(t-1)g0})^2 \label{c1b}\\
& && + \sum_{g \in \mathcal{G}_{t}}^{}\Big(\eta_{tg}^{+}\;r_{tg}^{+} + \eta_{tg}^{-}\;r_{tg}^{-}\label{c1c}\\
& && \quad\quad\quad\quad + \mu_{tg}^{+}\;w_{tg}^{+}+\mu_{tg}^{-}\;w_{tg}^{-} \Big)\Bigg)
\label{c1d}\\[7pt]
&\text{s.t.~~~} && \dbf_{tc}+\Bbf_{tc}\; \boldsymbol{\theta}_{tc} =\Cbf_{tc}^{\top}\; \pbf_{tc},  && \hspace{-1cm}\forall t, \forall c\label{eq:power_balance_cons}\\[5pt]
& && \text{where}\nonumber\\
& &&\pbf_{tc}\triangleq[
p_{t1c},\;p_{t2c},\;\ldots,\;p_{t|\Gcal_{tc}|c}
]^\top\nonumber\\[5pt]
& &&\lvert \vec{\Bbf}_{tc} \;\boldsymbol{\theta}_{tc} + \fbf^{\mathrm{shift}}_{tc}\rvert\leq \fbf^{\mathrm{max}}_{tc}, && \hspace{-1.9cm}\forall t, \forall c\label{eq:thermal_cons}\\[5pt]
& &&{p}_{gc}^{\mathrm{min}}\; \leq  p_{tgc} \leq {p}_{gc}^{\mathrm{max}}, && \hspace{-1.9cm}\forall t, \forall g, \forall c\!\label{eq:capacity_cons}\\[4pt]
& &&0 \leq w^{+}_{tg} \leq {w}_{tg}^{\mathrm{max}}, && \hspace{-1.9cm} \forall t, \forall g\label{eq:lframp_cons_3}\\[4pt]
& &&0 \leq w^{-}_{tg} \leq {w}_{tg}^{\mathrm{min}}, && \hspace{-1.9cm} \forall t, \forall g \label{eq:lframp_cons_4}\\[4pt]
& &&p_{tg0} - p_{(t-1)g0} \leq w^{+}_{(t-1)g}, && \hspace{-1.9cm} \forall t, \forall g\label{eq:lframp_cons_1}\\[4pt]
& &&p_{(t-1)g0} - p_{tg0} \leq w^{-}_{(t-1)g}, && \hspace{-1.9cm} \forall t, \forall g\label{eq:lframp_cons_2}\\[4pt]
& &&0 \leq r_{tg}^{+} \leq {r}_{tg}^{\mathrm{max}}, && \hspace{-1.9cm} \forall t, \forall g\label{eq:cont_res_cons_3}\\[4pt]
& &&0 \leq r_{tg}^{-} \leq {r}_{tg}^{\mathrm{min}}, && \hspace{-1.9cm} \forall t, \forall g\label{eq:cont_res_cons_4}\\[4pt]
& &&p_{tgc} - p_{tg0} \leq r_{tg}^{+}, && \hspace{-1.9cm} \forall t, \forall g, \forall c \neq 0\label{eq:cont_res_cons_1}\\[4pt]
& &&p_{tg0} - p_{tgc} \leq r_{tg}^{+}, && \hspace{-1.9cm} \forall t, \forall g, \forall c\neq 0\label{eq:cont_res_cons_2}\\[4pt]
& &&\!\!\!\!-{\Delta}_{g}^{\mathrm{min}} \leq p_{tgc} - p_{tg0} \leq {\Delta}_{g}^{\mathrm{max}}, && \hspace{-1.9cm} \forall t, \forall g, \forall c\neq 0\label{eq:cont_res_cons_5}
\end{align}
\end{subequations}
\endgroup

Objective \cref{c1a} represents the operation cost with quadratic coefficient $\alpha^{\mathrm{sqr}}_{tg}$, linear coefficient $\alpha^{\mathrm{lin}}_{tg}$ and fixed cost $\zeta_{tg}$. \cref{c1b} represents the quadratic
load-following ramp “wear and tear” cost with coefficient $\kappa_{g}$. Furthermore, \cref{c1c} accounts for the cost of contingency with coefficients $\eta_{tg}^{+}$ and $\eta_{tg}^{-}$. Expression \cref{c1d} represents cost of load-following ramping reserves with coefficients $\mu_{tg}^{+}$ and $\mu_{tg}^{-}$.
Constraint \cref{eq:power_balance_cons} imposes the Direct Current (DC) power balance constraints. Constraint \cref{eq:thermal_cons} restricts the flow of power by the vector of line thermal limits $\fbf^{\mathrm{max}}_{tc}\in  \Rbb^{|\Ecal_{tc}|}$, where $\fbf^{\mathrm{shift}}_{tc}$ accounts for the effect of transformers and phase shifters. Constraint \cref{eq:capacity_cons} imposes upper and lower limits \{${p}_{gc}^{\mathrm{min}}$ , ${p}_{gc}^{\mathrm{max}}$\} on each generator. Per-period ramp limits are imposed on the units with constraints \cref{eq:lframp_cons_1,eq:lframp_cons_2,eq:lframp_cons_3,eq:lframp_cons_4}. Constraints \cref{eq:cont_res_cons_3,eq:cont_res_cons_4,eq:cont_res_cons_1,eq:cont_res_cons_2} impose upward/downward limits \{${r}_{tg}^{\mathrm{max}}$ , ${r}_{tg}^{\mathrm{min}}$\} on post-contingency dispatch quantities, respectively. Additionally, constraint \cref{eq:cont_res_cons_5} enforces limits on downward and upward transitions from base to post-contingency state.

\subsection{Multi-Step Cascaded Optimization Problem considering Hurricane Progression}%\label{formulation}
% More details about the optimization problem formulation is given as follows:
% \subsection{Objective Function}
% The objective function can have multiple forms, which are given as follows:
% \begin{enumerate}
%     \item cost function: considering the generation production cost, which could include fuel cost, variable O\&M cost, cost of contingency and load following ramping reserve, potential contract violation cost, and load-following ramp cost, see ref \cite{21QEMR}
%     \item load impact: considering the aggregated metric of load impact, which is a function of bus voltage and served load, the smaller, the better. See ref \cite{osti_1771798}
%     \item load to be served
% \end{enumerate}

% \subsection{Constraints}
% The following types of contraints are applied in this problem:
% \begin{enumerate}
%     %\item integrality constraint
%     \item unit capacity constraint
%     %\item minimum up/down time constraint
%     \item ramp constraints
%     \item post-contingency reserve constraint
%     \item DC network constraint
% \end{enumerate}

% \subsection{multi-period formulation based on hurricane}
A hurricane can be divided into multiple steps considering its time progression. In this case, the resulted power flow case from previous step of existing Dynamic Contingency Analysis Toolbox (DCAT) simulation can be utilized in next step, as a new operation basecase. Then contingency list will be updated for individual step, as hurricane will continue to move through the power grid spanning over large geographical areas.

\begin{figure*}
  \begin{center}
    \includegraphics[scale=0.6]{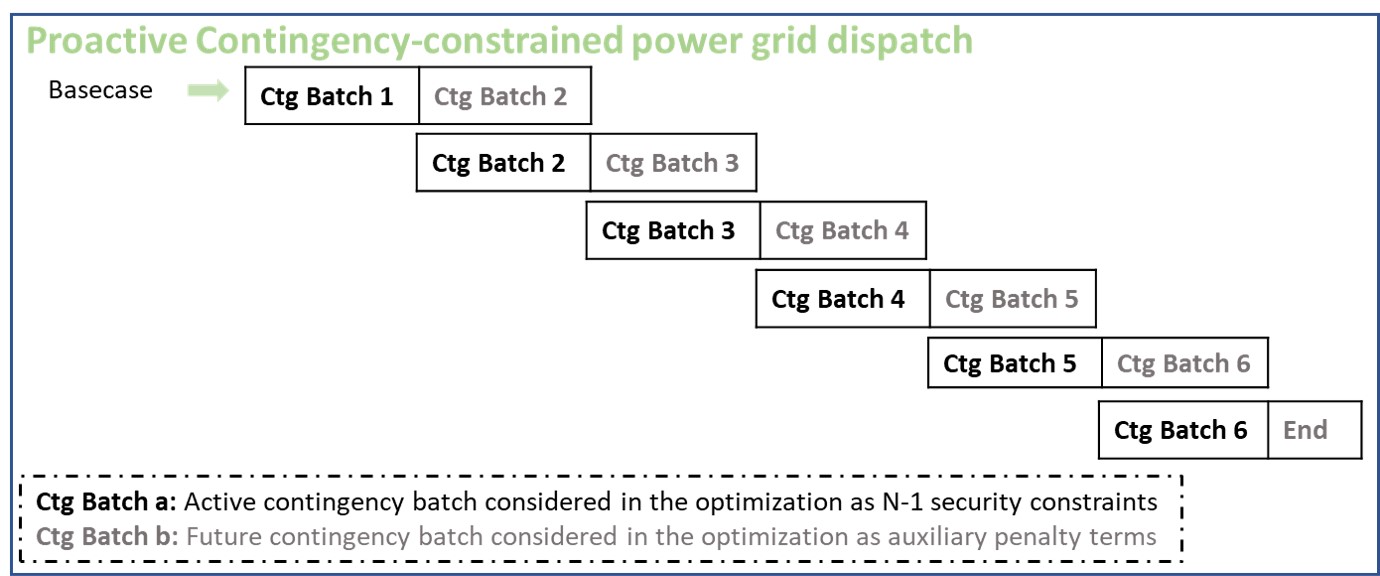}
  \end{center}
  \caption{An illustration of multi-step cascaded optimization considering hurricane progression.}
    \label{Fig:ctg_illustration}
\end{figure*}

It should be noted that with the predictive intelligence derived from hurricane modeling, one potential idea would be for one hurricane step, not only using the contingencies in the current step list, but also imposing further constraints on transmission line yet to-be-impacted in next step, i.e., more conservative constraints of line flow ratings (100\% to 70\%); therefore, the resulted contingency-constrained SCOPF problem is further extended with proactive posturing capability throughout the  hurricane life cycle. Fig. \ref{Fig:ctg_illustration} provides an illustrative example with 6 steps representing one historical hurricane.% More details about the optimization problem formulation is given in the following subsections.

\section{Convexification of the Proactive SCOPF Problem}\label{convexification}
In order to efficiently solve the the optimization problem, we propose SDP relaxation which lead to a computationally-tractable algorithm following \cite{21QEMR}.
\subsection{Lifting}
The first stage of convex relaxation is to lift the problem to a higher dimensional space. This is done by introducing the auxiliary variables $\mathbf{O}_{tgc},  \mathbf{h}_{tgc}\in\Rbb^{|\Tcal|\times|\Gcal|\times|\Ccal|}$ accounting for the monomials $p^2_{tgc}$ and $p_{tgc}p_{(t-1)gc}$ respectively. Using the defined variables, the objective function \cref{c1a} can be cast linearly as \cref{eq:obja_lift}. Likewise the objective function \cref{c1b} can be cast linearly as \cref{eq:objb_lift}. Furthermore, in order to enforce the relationship between the auxiliary variables and the corresponding monomials, we strengthen the proposed convex relaxation via conic constraint \cref{eq:conic_cons_1}. The relationship between the auxiliary variable $\mathbf{O}_{tgc}$ and monomial $p^2_{tgc}$ is relaxed to the SOCP \cref{eq:ooo1}. %T

\subsection{SDP relaxation}
The resulting SDP relaxation of the proactive SCOPF problem is given by:

\begin{subequations}\label{convex_relaxation}
\begin{align}
&\underset{}{\text{min~~~}} && \!\! \sum_{t \in \mathcal{T}}^{} \Bigg(
\sum_{c \in \mathcal{C}_{t}}^{}
\sum_{g \in \mathcal{G}_{tc}}^{}
\alpha^{\mathrm{sqr}}_{tg}\; \mathbf{O}_{tgc} +\alpha^{\mathrm{lin}}_{tg}\; p_{tgc} \,+ \zeta_{tg}\label{eq:obja_lift} \\
& && + \sum_{c \in 0}^{}
\sum_{g \in \mathcal{G}_{tc}}^{}\!\!\! \kappa_{g}\!\times\!(\mathbf{O}_{tg0} \;+\mathbf{O}_{(t-1)g0} - 2\times \mathbf{h}_{tgc})\!\!\!\!\!\label{eq:objb_lift}\\
& && + \sum_{g \in \mathcal{G}_{t}}^{}\Big(\eta_{tg}^{+}\;r_{tg}^{+} + \eta_{tg}^{-}\;r_{tg}^{-}\label{eq:objc_lift}\\
& && \quad\quad\quad\quad + \mu_{tg}^{+}\;w_{tg}^{+}+\mu_{tg}^{-}\;w_{tg}^{-} \Big)\Bigg)\label{eq:objd_lift}\\[5pt]
&\text{s.t.~~~} &&\mathbf{W}\quad\succeq\quad\mathbf{w}\mathbf{w}^\top,&&\hspace{-3.5cm}\forall tgc\in\mathcal{T}\times\mathcal{G}\times\mathcal{C}\label{eq:conic_cons_1}\\[5pt]
& && \mathbf{W} =    \begin{bmatrix*}[l]
  O_{(t-1)gsc}\! & \ast\ \\[5pt]
  h_{tgc}\! &\!\! O_{tgc}\! \end{bmatrix*}\!, \mathbf{w} = \begin{bmatrix*}[l]
  p_{(t-1)gc}\!\\ p_{tgc}\! \end{bmatrix*}, \nonumber\\[5pt]
  & && O_{tgc} \geq p^2_{tgc} \;,\; O_{tgc}\geq 0 &&\hspace{-2cm} \forall t, \forall g, \forall c\label{eq:ooo1}\\[5pt]
  & && O_{tgc}+ {p}_{gc}^{\mathrm{min}}\,{p}_{gc}^{\mathrm{max}}\, \leq ({p}_{gc}^{\mathrm{min}} + {p}_{gc}^{\mathrm{max}})\;p_{tgc}\nonumber\\
  & && && \hspace{-2cm} \forall t, \forall g, \forall c\\[5pt]
  & &&\text{\cref{eq:power_balance_cons,eq:capacity_cons}}&& \hspace{-2cm} \forall t, \forall g, \forall c\\[5pt]
 & &&\text{\cref{eq:lframp_cons_1,eq:lframp_cons_2,eq:lframp_cons_3,eq:lframp_cons_4}}&& \hspace{-2cm} \forall t, \forall g\\[5pt]
 & &&\text{\cref{eq:cont_res_cons_1,eq:cont_res_cons_2,eq:cont_res_cons_3,eq:cont_res_cons_4,eq:cont_res_cons_5}}&& \hspace{-2cm} \forall t, \forall g, \forall c
\end{align}
\end{subequations}

The above relaxation is exact if equally holds for \cref{eq:conic_cons_1} and \cref{eq:ooo1}, making the solution to the optimization problem globally optimal. When the relaxation is not exact, we use the following measure to evaluate the closeness to global optimality:
\begin{equation}
\text{Optimality Gap}\;\% = 100 \times \frac{f_r - f^{opt}}{f_r}
\end{equation}
where $f_r$ and $f^{opt}$ are the optimal objective values for convex relaxation \eqref{convex_relaxation} and the globally optimal solution to the original problem \eqref{initial_problem}, respectively. In summary, if the SDP relaxation has a rank-1 solution, then a feasible solution of proactive SCOPF together with a global optimality guarantee can be computed.

\section{Numerical Results}\label{numerical_results}
To evaluate and validate our proposed methodology, a real-world power grid model is utilized considering a single historical hurricane event. The power system model considered is a 1263-Bus,  1269-Branch network\cite{osti_1771798}. Based on historical hurricane event data, one realization of the hurricane was derived considering wind speed variations. Moreover, within this realization the hurricane process is divided into six consecutive steps; each step is a set of power system contingencies, such as credible transmission line, bus and generator outages along the hurricane trajectory shown in Table \ref{tab:contingency data}. Detailed \textcolor{black}{network topology, simulation} examples and illustrations can be found in \textit{Section 5, Simulation Results} in \cite{osti_1771798}.

Simulations are performed on a laptop with an Intel Core i5 2.3 GHz CPU and 8 GB RAM using MATLAB R2020b. The open-source solver SDPT3 is used to solve the semidefinite program through CVX v2.1\cite{cvx}. The  objective function for each contingency batch is the sum of operation cost over time scale of 15 minutes as each contingency (Ctg) batch from DCAT has the same time scale. For each step in the hurricane progression, grid data obtained through DCAT is fed as a basecase to the optimization problem. The output of the optimization is then analyzed. This multi-step cascaded optimization setup is illustrated in Figure \ref{Fig:ctg_illustration}.

\begin{table}[hbt!]
\begin{center}\scriptsize
\caption{Contingency Data and Operation Cost}
\label{tab:short-epoch}
\begin{tabular}{@{\,}c|c|c|c|c}\hline
Contingency & \# of Lost & \# of Lost & \# of Lost & Operation\\
        Batches         & Buses  & Branches  & Generators  & Cost (\$)\\\hline\hline
Ctg Batch 1 & 0  & 0 & 0 & 99,638 \\
Ctg Batch 2 & 0  & 1 & 0 & 114,248 \\
Ctg Batch 3 & 90 & 114 &  2 & --\\
Ctg Batch 4 &  268 & 318 & 4 & 1,537,810 \\
Ctg Batch 5 &  376 & 447 & 5 & --\\
Ctg Batch 6 & 401  & 481 & 5 & --\\\hline
\hline
\end{tabular}
\label{tab:contingency data}
\end{center}
\end{table}

\subsection{Operation Cost}
In the simulations, the optimizer found an optimal solution to the SDP relaxation after computation time of 2 minutes 46 seconds. It was observed that the cost of operating the grid during hurricane-triggered contingencies gradually increases as the hurricane progresses through the grid network and causes loss to network elements. This result is shown in Table \ref{tab:contingency data}, operation cost gradually increases from a basecase cost of \$99,638 to \$1,537,810 in Ctg Batch 4. This can be attributed to increased utilization of network elements as a result of the loss of buses, branches and generators in the network making the electric grid more and more expensive to operate. In Ctg Batch 3, 5, and 6 the model becomes infeasible due to the grids inability to survive outage to several damaged elements.

\subsection{Proactive Mitigation Measure}
Hurricane impacts which lead to fatal damage to grid elements such as lines, buses, generators, transformer etc. may be mitigated by applying proactive measures to adjust their operation during hurricane progression while still maintaining reliability to serve as many customers as possible. To test proactive measures, we apply tighter line flow constraints to Ctg Batch branches to power flow cases generated through DCAT's time domain simulation. In the experiments, line flows on branches in the hurricane's path are proactively limited to operate at 70\% of normal operation limits.

\begin{figure}[h!]
	\centering
	\includegraphics[width=0.5\textwidth]{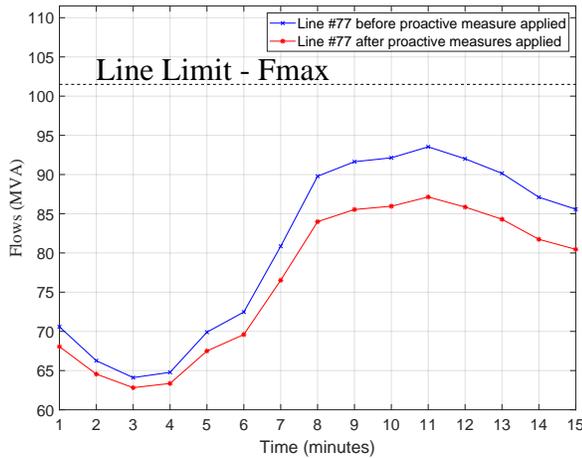}
	\caption{Line MVA flows on line \# 77 before and after applying proactive mitigation to penalize the flow on selected lines in the hurricane path}
	\label{fig:lineflow}
\end{figure}

In Figure \ref{fig:lineflow}, we observe that line flows over branch \#77 reduces by a maximum of 7\% after we applied a proactive mitigation strategy of tightening branch flow constraints without violating other constraints or causing optimization infeasibility. % With a less conservative approach,
 We then further tighten the line flow limits below 70\% until we reach infeasibility of the model for the purpose of analysis; additional  runs with progressively tighter branch flow constraints indicated that the model becomes infeasible when line flow limits fall below 40\% of normal operation while still meeting the system load demand.

\section{Conclusions}\label{conclusions}
The paper aims to develop interactive cross-domain data analytics based on hurricane modeling, power system optimization and model-based simulations. The temporal relations among individual steps (groups of time period) within historical hurricane event have been explored and transformed into explicit optimization constraints, and further incorporated into SCOPF problem to identify proactive posturing of power system elements. Variations of the operation cost among different hurricane steps (contingency batches) indicates the applicability of such optimization formulation, and the improvement in targeted credible line contingencies shows promising  improvements by such proactive dispatch. Future work includes scalability evaluation with large volume of DCAT simulation outputs based on different base cases and synthesized hurricane events, and expanded validation considering multiple resilience metrics\cite{osti_1771798}.
%Natural disasters such as hurricanes are challenging the operation and control of U.S. power grid more frequently in the past decade, and it is crucial to develop proactive strategies to assist grid operators for better emergency response and minimized electricity service interruptions. In this paper, we propose a proactive posturing of power system elements, including genration, transmission and load, and formulated a security constrainted optimal power flow (SCOPF) that informed by cross-domain hurricane modeling and its impacts on the grid elements. Simulation results based on real-world power grid and historical hurricane event show improvements in decreased xxx and increased yyy, as well as zzz.

%\section*{Acknowledgment}

% The preferred spelling of the word ``acknowledgment'' in America is without
% an ``e'' after the ``g''. Avoid the stilted expression ``one of us (R. B.
% G.) thanks $\ldots$''. Instead, try ``R. B. G. thanks$\ldots$''. Put sponsor
% acknowledgments in the unnumbered footnote on the first page.

\bibliographystyle{IEEEtran}
\bibliography{IEEEabrv,reference}

\end{document}